\documentclass[letterpaper, 10 pt, journal]{IEEEtran} 

\IEEEoverridecommandlockouts 

\usepackage{graphicx} 
\usepackage{amsmath} 
\usepackage{amssymb} 
\usepackage{amsfonts,amsthm} 
\usepackage[usenames,dvipsnames]{xcolor}

\usepackage{enumitem}

\title{Novel Stability Conditions for Nonlinear Monotone Systems and Consensus in Multi-Agent Networks}

\author{D. Deplano, \IEEEmembership{Member, IEEE}, M. Franceschelli, \IEEEmembership{Senior, IEEE}, and A. Giua, \IEEEmembership{Fellow, IEEE}
\thanks{
This work was partially supported by Region Sardinia (RAS) with project MOSIMA, RASSR05871, FSC 2014-2020, Annualità 2017, Area Tematica 3, Linea d'Azione 3.1. and by the Fondazione Banco di Sardegna with the grant \virg{Formal Methods and Technologies for the Future of Energy Systems}, cup F72F20000350007.
}
\thanks{D. Deplano, M. Franceschelli and A. Giua are with DIEE, University of Cagliari, 09123 Cagliari, Italy.
Emails: {\tt \{diego.deplano,mauro.franceschelli,giua\}@unica.it}}
\thanks{\textbf{\color{red}Preprint submitted to \virg{IEEE Transactions on Automatic Control} on the 3rd of December 2021, currently under review.}}
}

\newcommand{\virg}[1]{``#1"}
\newcommand{\abs}[1]{{\left\vert #1 \right\vert}}
\newcommand{\norm}[1]{{\left\vert\kern-0.25ex\left\vert #1\right\vert\kern-0.25ex\right\vert}}
\newcommand{\norminf}[1]{{\left\vert\kern-0.25ex\left\vert #1\right\vert\kern-0.25ex\right\vert}_{\infty}}

\newcommand{\rea }{\mathbb{R}}
\newcommand{\nat}{\mathbb{N}}
\newcommand{\tea }{\mathbb{T}}

\newcommand{\clt}{\tag*{$\blacksquare$}}

\newcommand{\Xc}{\mathcal{X}}
\newcommand{\Fc}{\mathcal{F}}
\newcommand{\Tc}{\mathcal{T}}
\newcommand{\Gc}{\mathcal{G}}
\newcommand{\Nc}{\mathcal{N}}
\newcommand{\bone}{\mathbf{1}}
\newcommand{\bzero}{\mathbf{0}}

\newtheorem{assum}{Assumption}
\newtheorem{thm}{Theorem}
\newtheorem{rem}{Remark}
\newtheorem{cor}{Corollary}
\newtheorem{lem}{Lemma}
\newtheorem{defn}{Definition}

\usepackage[T1]{fontenc}

\usepackage{csquotes}

\makeatletter
\let\c@author\relax
\makeatother
\usepackage[citestyle=numeric-comp,
 style= ieee,
 sorting=none,
 giveninits=true,
 sortcites=true]{biblatex}
\AtEveryBibitem{
	\clearfield{issn}
	\clearfield{isbn}
	\clearfield{archivePrefix}
	\clearfield{arxivId}
	\clearfield{pmid}
	\clearfield{doi}
	\clearfield{eprint}
	\clearfield{note}
	\clearfield{month}
	\ifentrytype{online}{}{\ifentrytype{misc}{}{\clearfield{url}}}
	\ifentrytype{book}{\clearfield{doi}}{}
}

\begin{document}
\maketitle

\begin{abstract}
In this work, we characterize a class of nonlinear monotone dynamical systems that have a certain translation invariance property which goes by the name of plus-homogeneity; usually called \virg{topical} systems. Such systems need not be asymptotically stable, since they are merely nonexpansive but not contractive.
Thus, we introduce a stricter version of monotonicity, termed \virg{type-K} in honor of Kamke, and we prove the asymptotic stability of the equilibrium points, as well as the convergence of all trajectories to such equilibria for type-K monotone and plus-homogeneous systems: we call them \virg{K-topical}.

Since topical maps are the natural nonlinear counterpart of linear maps defined by row-stochastic matrices, which are a cornerstone in the convergence analysis of linear multi-agent systems (MASs), we exploit our results for solving the consensus problem over nonlinear K-topical MASs.
We first provide necessary and sufficient conditions on the local interaction rules of the agents ensuring the K-topicality of a MAS. 
Then, we prove that the agents achieve consensus asymptotically if the graph describing their interactions contains a globally reachable node.

Finally, several examples for continuous-time and discrete-time systems are discussed to corroborate the enforceability of our results in different applications.
\end{abstract}


\section{Introduction}

Dynamical systems whose trajectories preserve a partial order have represented a fruitful topic of research in numerous fields: such systems are usually called \emph{monotone}~\cite{Smith08}. 
Among all particular classes of monotone systems, this paper considers those ruled by \emph{topical} maps, which are the nonlinear counterpart of linear maps defined by \emph{row-stochastic matrices}~\cite{Horn12}.

Following Gunawardena and Keane~\cite{Gunawardena1995}, we denote by the name \emph{topical} those systems whose solutions or \emph{flows} $\varphi$ satisfy
\begin{align}\label{eq:mono}
 x \leq z &\Rightarrow \varphi(t,x)\leq \varphi(t,z),&& \forall x,z \in \Xc, \\
 \varphi(t, x + \alpha\bone) &= \varphi(t,x) + \alpha\bone,&&  \forall x\in \rea^n,\forall \alpha\in\rea \label{eq:p-homo},
\end{align}
at any time $t\geq 0$, where $x,y$ denote initial conditions in the state space $\Xc\subseteq \rea^n$.
We refer to the property in eq.~\eqref{eq:mono} as \emph{monotonicity} and to the property in eq.~\eqref{eq:p-homo} as \emph{plus-homogeneity}. Topical dynamical systems have been a subject of interest of both monotone dynamical systems theory~\cite{Angeli08}, where plus-homogeneity is referred to as \virg{translation invariance}, and nonlinear Perron–Frobenius theory~\cite{LemmensNussbaum2012}, where monotonicity is referred to as \virg{order-preservation}.
In this paper, we introduce a stricter variation of monotonicity, called \emph{type-K monotonicity} in honor of Kamke, which can be seen as an important bridging link between these two theories, as it is discussed in Section~\ref{sec:lit_rev}.
Consequently, we call \emph{K-topical} those systems being type-K monotone and plus-homogeneous

\subsection{Main contributions}
The main goal of this paper is to give a self-contained introduction to smooth K-topical systems both in continuous-time, where \virg{smooth} denotes the continuous differentiability of the vector field, and in discrete-time, where \virg{smooth} denotes the continuous differentiability of the map.
Within this goal, our first main result is the following:
\begin{itemize}
    \item Trajectories of smooth K-topical systems are proved to asymptotically converge toward an equilibrium point, if any exists (see~Theorem~\ref{th:convergence}).
\end{itemize}
A further contribution is the derivation of necessary and sufficient conditions for type-K monotonicity:
\begin{itemize}
    \item A smooth continuous-time system is type-K monotone if and only if its Jacobian matrix is Metzler everywhere (see~Corollary~\ref{cor:kamkeKcond});
    \item A smooth discrete-time system is type-K monotone if and only if its Jacobian matrix is Metzler everywhere with a strictly positive diagonal almost everywhere (see~Theorem~\ref{thm:kamke_like_cond}).
\end{itemize}
A knowledgeable reader may recognize the similarity of these conditions to the well-known \emph{Kamke condition} for continuous-time system \cite{Kamke32, Coppel65, Smith08}. Indeed, a remarkable result is that the Kamke condition is necessary and sufficient not only for monotonicity of smooth systems (see~Corollary~\ref{thm:kamke_cond}), but also for type-K monotonicity:
\begin{itemize}
    \item Monotone systems in continuous-time whose vector field is continuously differentiable are type-K monotone (see~Theorem~\ref{th:monotoneK}).
\end{itemize}
A second goal consists in exploiting the convergence result and the characterization of K-topical systems presented above to solve the consensus problem in K-topical Multi-Agent Systems (MASs). 
Most of the results for achieving consensus in linear MAS have been derived by considering row-stochastic matrices, for which the celebrated Perron-Frobenius theory provides a thorough spectral characterization~\cite{Wolfowitz63, Jadbabaie03, Bullo18}.
Since topical maps generalize linear maps defined by row-stochastic matrices, our results lay the groundwork for a systematic analysis of general MAS with nonlinear interaction rules among agents.
Within this goal is our last result:
\begin{itemize}
    \item A K-topical MAS achieves consensus asymptotically if the origin is an equilibrium point and if the graph contains a globally reachable node (see~Theorems~\ref{th:consensus_CT}-\ref{th:consensus_DT}).
\end{itemize}

\subsection{Literature review }\label{sec:lit_rev}
In the theory of monotone dynamical systems, emphasis is put on the class of continuous-time systems being strongly monotone~\cite{Angeli08}, i.e., whose flows possess the following property:
$$
x \lneq z \Rightarrow \varphi(t,x)< \varphi(t,z),\qquad \forall x,z \in \rea^n.
$$
Pioneering work in this field was done by Hirsch, who first showed that if solutions of continuous-time strongly monotone dynamical systems exist and are bounded, then they converge to a set of equilibrium points~\cite{Hirsch88}. On the other hand, for discrete-time strongly monotone dynamical systems, Polavcik showed that their iterative behavior converges to periodic points under appropriate additional conditions~\cite{Polavcik92}. An extensive overview of these results was given by Hirsch and Smith~\cite{Smith08, Hirsch06}. Remarkably, generic convergence to equilibria can be made global, as in the case of contractive systems with a unique equilibrium point~\cite{Banach22}.

In contrast, in nonlinear Perron–Frobenius theory one usually considers discrete-time dynamical systems that are only monotone~\cite{LemmensNussbaum2012}. However, the relaxation of the assumption of strong monotonicity makes unenforceable most of the theory of monotone systems which then requires some additional assumptions. An interesting branch of research has focused on \emph{topical systems} which possess the plus-homogeneity property in eq.~\eqref{eq:p-homo}~\cite{Hammouri99,Akian12,Doban14,Dong15,Feyzmahdavian13,Sanchez19}
, as well as its extension to the multi-homogeneous systems~\cite{Friedland13,Gautier16}
: a unified framework has been recently provided by Gautier et. al. in~\cite{Gautier2019}. The pioneering work of Nussbaum~\cite{Nussbaum1988} showed that topical systems are nonexpansive under the sup-norm, contrary to the strong monotonicity assumption which causes the system to be contractive, thus ensuring the convergence of all trajectories to an equilibrium point by a direct application of the Banach fixed point theorem~\cite{Banach22}. Indeed, when the system is merely nonexpansive, such a nice global convergence result is lost and one can only show that the trajectories converge to periodic points and thus not necessarily to an equilibrium point.
Nussbaum has also shown that the primitiveness of the Jacobian matrix is a sufficient condition ensuring the convergence of a differentiable discrete-time system to its positive eigenvector; this result has recently been generalized to multi-homogeneous systems in~\cite{Gautier2019}.

The control community has recetly recognized the importance of bridging the two above-mentioned approaches.
Angeli and Sontag were the first 
to consider topical systems~\cite{Angeli06, Angeli08}. In particular, they have proved that every solution of continuous-time topical systems possessing the strong monotonicity property converges to an equilibrium point if the trajectory is bounded. If one wishes to get a global convergence result only assuming that the dynamical system is monotone without a stronger assumption, one meets several difficulties when applying any known methods used in the strongly monotone case.
Afterward, Hu and Jiang provided a similar result for the restricted class of time-periodic systems while getting rid of the strong monotonicity assumption~\cite{Hu10}. Their proof methodology is interesting: they provide a global convergence result of discrete-time systems ruled by the Poincaré map associated with a time-periodic topical system, which is, in turn, a topical system possessing the property of \emph{type-K monotonicity}.
The type-K monotonicity property, which encompasses strong monotonicity, has been proposed for the first time by Jiang in~\cite{Jiang1996}, and it has been recently exploited in the context of multi-agent systems by us in~\cite{Deplano18, Deplano20}.

There are many authors currently investigating the consensus problem over nonlinear monotone networks and systems, which sometimes intrinsically possess the plus-homogeneity property. Among them, Manfredi and Angeli have studied the case of monotone networks with unilateral interactions \cite{Manfredi17}. Como and Lovisari have considered monotone dynamical flow networks \cite{Como16, Lovisari14}, a topic of interest for Coogan and Arcak as well \cite{Coogan16}. In particular, Coogan has recently presented a tutorial paper on mixed monotonicity, which extends the usual notion of monotonicity \cite{Coogan20}. Worthy of mention is also the line of research on eventually monotone systems pursued by Altafini and Mauroy \cite{Sootla18, Altafini14}, as well as the framework of differentially positive systems drawn up by Forni and Sepulchre \cite{Forni15}, and also the operator-theoretic perspective adopted by Belgioioso and Grammatico \cite{Belgioioso18}. For insights on new advances and applications of monotone systems, we refer the interested reader to the recent work of Smith~\cite{Smith17}.

\subsection{Structure of the paper}
In Section~\ref{sec:background} we introduce the notation of the paper along with some required preliminaries. In Section~\ref{sec:dynamicalSystems} we provide a global convergence result for K-topical dynamical systems. In Section~\ref{sec:multiAgent} we consider K-topical multi-agent systems and provide additional results regarding the consensus problem. In Section~\ref{sec:examples} we discuss several examples to corroborate the applicability of our results. Finally, in Section~\ref{sec:conclusions} we give our final remarks and outline potential future directions.

\section{Notation and preliminaries} \label{sec:background}

The set of real and integer numbers are
denoted by $\rea$ and $\mathbb{Z}$, while their restriction to nonnegative values are denoted with $\rea_{\geq 0}$ and $\nat$, respectively.
Matrices are denoted by uppercase letters, vectors and scalars are denoted by lowercase letters, while sets are denoted by uppercase calligraphic letters.
We denote by $\bzero_n$ and $\bone_n$ the vector of zeros and ones of dimension $n$, respectively. The identity matrix of dimension $n$ is denoted by $I_n$. If clear from the context, subscripts are omitted.

\subsection{Dynamical systems}
 
 We consider \emph{autonomous dynamical systems} with an euclidean \emph{state space} $\Xc\subseteq \rea^n$ and denote the \emph{state} of the system at a generic time $t$ by $x(t)\in \Xc$.
 \begin{assum}
  The domain $\mathcal{X}\subseteq \mathbb{R}^n$ is assumed to be open and convex, i.e., $(1 - \alpha)x + \alpha y \in \Xc $ for all $x,y\in \Xc$.

\end{assum}

When time is a continuous variable, $t\in \rea$, the system is described by a set of ordinary differential equations arising from,
\begin{equation*}
    \dot{x}(t) = f(x(t)),\qquad t\in\mathbb{R}.
\end{equation*}

When time is a discrete variable, $k\in \nat$, the system is described by a set of difference equations,
\begin{equation*}
    x(k+1) = f(x(k)),\qquad k\in\nat.
\end{equation*}

Function $f$ determines the evolution of the state in time: in continuous-time, $f:\Xc\rightarrow \rea^n$ is a vector field; in discrete-time, $f:\Xc\rightarrow\Xc$ is a map. We limit our study to smooth systems, which are systems satisfying the following standing assumption.

\begin{assum}
In both frameworks, function $f$ is assumed to be of class $C^1$, i.e., $f$ is continuously differentiable.
\end{assum}

 Since we consider both continuous-time and discrete-time systems, it is convenient to describe a dynamical system in terms of its flow. Such description applies to both frameworks and allows us to use a general uniform notation throughout the paper. To this aim, we denote the time domain $\tea$ which has to be intended as follows:
 \begin{itemize}
     \item $\tea  = \rea$ for continuous-time systems;
     \item $\tea  = \nat$ for discrete-time systems;
 \end{itemize}

To emphasize the dependence of the evolution $x(t)$ on the initial state $x(0)=\xi$, we denote the corresponding evolution by $\varphi(t,\xi)$, i.e.,
$$
\varphi(t,\xi) = x(t),\quad \text{if}\quad  x(0)=\xi.
$$

The map $\varphi(t,\xi):\mathbb{T}\times \Xc \rightarrow \Xc$ is called the \emph{flow} of the system at time $t\in\tea$ starting at $\xi$.
The sequence of all consecutive states of the system is called the \emph{trajectory} of the system, and it is denoted by $\Tc(\xi)=(\varphi(t,\xi))_{t\geq 0}$.
A trajectory $\Tc(\xi)$ is said to be \emph{bounded} if there exist $\ell,u\in\Xc$ such that for all $x\in\Tc(\xi)$ it holds $\ell\leq x\leq u$; otherwise it is said to be \emph{unbounded}.

A point $\xi\in\Xc$ is called \emph{periodic} if there exists a positive $T$ such that $\varphi(T,\xi)=\xi$. The minimal such $T$ is called the period of $x$. If the relation holds for any $T\in \mathbb{R}_{\geq 0}$, we call $\xi$ an \emph{equilibrium point}. We denote by ${\Fc(\varphi)=\{\xi\in \Xc:\varphi(t,\xi)=\xi,\forall t\in\tea\}}$, the set of equilibrium points, or simply $\Fc$ when clear from the context.
An equilibrium point $x_e\in \Fc(\varphi)$ is said to be \emph{stable} if for every $\varepsilon >0$ there is $\delta > 0$ such that
$\norm{\xi-x_e}<\delta$ implies ${\norm{\varphi(t,\xi)-x_e}<\varepsilon}$ for any $\xi\in\Xc$ and $t\in\tea$, where $\norm{\cdot}$ denotes the norm of a vector.

\subsection{Multi-agent systems}

We consider Multi-Agent Systems (MASs) wherein the ${n\in\nat}$ agents are modeled as autonomous dynamical systems with scalar state ${x_i(t)\in\rea }$, for $i=1,\ldots,n$.

The interconnections among the agents are given by a graph $\Gc=(\mathcal{V},\mathcal{E})$ where $\mathcal{V}=\{1,\ldots,n\}$ is the set of nodes representing the agents and $\mathcal{E}\subseteq \mathcal{V} \times \mathcal{V}$ is a set of directed edges. A \emph{directed edge} $(i,j)\in \mathcal{E}$ exists if agent $i$ is influenced by agent $j$: in this case, agent $j$ is said to be a \emph{neighbor} of agent $i$. The set of neighbors of the ${i\text{-th node}}$ is denoted by $\Nc_i=\left\{j\in \mathcal{V}: (i,j)\in \mathcal{E}\right\}$. Each agent $i\in \mathcal{V}$ updates its own state according to a local interaction protocol, which, in continuous-time, takes one the form
$$
 \dot{x}_i(t) = f_i\left(x_i(t),x_j(t):\Nc_i\right), \qquad t\in \rea,
$$
and, in discrete-time, it takes the form
$$
 x_i(k+1) = f_i\left(x_i(k),x_j(k):\Nc_i\right),\qquad k\in \nat.
$$

A \emph{directed path} between two nodes $p$ and $q$ in a graph is a finite sequence of $m$ edges $e_k=(j_k,i_k)\in \mathcal{E}$ that joins node $p$ to node $q$, i.e., $j_1=p$, $i_m=q$ and $i_k=j_{k+1}$ for $k=1,\ldots,m-1$. The node $i$ is said to be \emph{reachable} from node $j$ if there exists a directed path from node $i$ to node $j$. A node is said to be \emph{globally reachable} if it is reachable from all nodes $j\in \mathcal{V}$.

A MAS is said to \emph{achieve consensus asymptotically} if the agents' states converge to the same constant value, called the \emph{consensus state}, i.e., there is $c\in\mathbb{R}$ such that
$$\lim_{t\rightarrow \infty}
x(t) = c\mathbf{1},\quad \text{ or }\quad 
\lim_{k\rightarrow \infty}
x(k) = c\mathbf{1},
$$
for any initial condition $x(0)\in\Xc$.

\subsection{ K-topical systems}\label{Intro:nonlinearPerron}

Consider the Euclidean space $\rea^n$ equipped with the standard partial order $\leq$ and let $\Xc \subseteq \rea^n$.
%
Dynamical systems in~${(\Xc, \leq)}$ whose flow preserves such order 
are referred to as \emph{order-preserving} or \emph{monotone} dynamical systems~\cite{Angeli03,Hirsch06,LemmensNussbaum2012}; we use the latter denomination. 
%
%
Next, we define a stricter notion of monotonicity termed \emph{type-K monotonicity}, which was introduced by us for dynamical systems in discrete-time~\cite{Deplano18, Deplano20}, while here it is given also for systems evolving in continuous-time.
%
%
\begin{defn}[{Monotonicity and type-K}]\label{def:type-k}
~
A dynamical system on $\Xc\in\rea^n$ is said to be \virg{monotone} if for any initial conditions $\xi_1,\xi_2\in \Xc$ the flow $\varphi:\tea\times\Xc\rightarrow \Xc$ satisfies
\begin{equation*}
\xi_1 \leq \xi_2 \Rightarrow \varphi(t,\xi_1)\leq \varphi(t,\xi_2),\quad  \forall t\in\tea,
\end{equation*}
and it is said to be \virg{type-K monotone} if, for all ${i=1,\ldots,n}$, it further satisfies
\begin{equation*}
\xi_{1,i} < \xi_{2,i} \Rightarrow \varphi_i(t,\xi_1)< \varphi_i(t,\xi_2), \quad \forall t\in\tea
\end{equation*}
where $\xi_{1,i}$, $\xi_{2,i}$ and $\varphi_i$ denote the $i$-th components. Correspondingly, the map $\varphi$ is said to be \virg{monotone} or \virg{type-K monotone}, respectively.
\end{defn}
%

%
%
%
We consider type-K monotone systems which are also invariant with respect to a rigid translation proportional to $\bone$. These systems are usually referred to as \emph{translation invariant} or \emph{plus-homogeneous}\footnote{The name \emph{plus-homogeneity} comes from the fact that the homogeneity is intended with respect to the addition operation, while simple \emph{homogeneity} is usually intended with respect to the multiplication operation, i.e., ${\varphi(t,\alpha \xi) = \alpha \varphi(t,\xi)}$, cfr.~\cite{LemmensNussbaum2012}} systems~\cite{Hirsch06,LemmensNussbaum2012}; we use the latter denomination.
%
%
%
%
\begin{defn}[{Plus-homogeneity}]\label{def:p-homo}
~
A dynamical system on $\Xc\in\rea^n$ is said to be \virg{plus-homogeneous} if the flow ${\varphi:\tea\times\Xc\rightarrow \Xc}$ satisfies
$$
\varphi(t,\xi+\alpha\bone) = \varphi(t,\xi)+\alpha \bone, \quad \forall\alpha \in\rea,\: t\in\tea
$$ for all initial conditions $\xi\in \Xc$. Correspondingly, the map $\varphi$ is said to be {plus-homogeneous}.
\end{defn}

Monotone systems satisfying also the plus-homogeneity property are known in the literature as \emph{topical} systems~\cite{Rubinov01,LemmensNussbaum2012,Mohebi14,Barsam16}.
Since we require the stricter type-K property, we next define the class of \emph{K-topical} systems.
\begin{defn}[{K-topicality}]\label{def: K-topicality}
~
A dynamical system on ${\Xc\subseteq \rea^n}$ is called \virg{K-topical} if it is type-K monotone and plus-homogeneous. Correspondingly, the map ${\varphi:\tea\times\Xc\rightarrow\Xc}$ is said to be K-topical.
\end{defn}

A nice feature of K-topical systems is that they are non-expansive w.r.t. the sup-norm; this property is widely known in the discrete-time framework \cite{Crandall80}, while in Lemma~\ref{lem:non-expansiveness} we prove it also for the continuous-time framework.
\begin{defn}[{Non-expansiveness}]\label{def:non-expansiveness}
~
A dynamical system on $\Xc\subseteq \rea^n$ is said to be \virg{non-expansive} w.r.t. a metric $d:\Xc\times\Xc\rightarrow \rea_{\geq 0}$ if the flow $\varphi$ satisfy
$$
d(\varphi(t,\xi_1),\varphi(t,\xi_2))\leq d(\xi_1,\xi_2),\quad \forall t\in\mathbb{T}$$
for all initial conditions $\xi_1,\xi_2\in\Xc$. Correspondingly, the map $\varphi:\tea\times\Xc\rightarrow \Xc$ is said to be non-expansive.
\end{defn}

\begin{lem} \label{lem:non-expansiveness} 
~
 K-topical systems on $\Xc\subseteq \rea^n$ are \virg{non-expansive} w.r.t. the {sup-metric} ${d_{\infty}:\Xc\times \Xc \rightarrow \rea_{\geq 0}}$ induced by the sup-norm, i.e.,
$$
d_{\infty}(\xi_1,\xi_2) = \norm{\xi_1-\xi_2}_{\infty},\quad \forall \xi_1,\xi_2\in\Xc.
$$
\end{lem}

\begin{proof}
For each fixed $t\in\tea$, we define a map ${\phi^t(x)=\varphi(t,x):\Xc\rightarrow\Xc}$. According to Crandall and Tartar~\cite[Proposition 2]{Crandall80}, each $\phi^t(\xi)$ is such that
$$
\norminf{\phi^t(\xi_1)-\phi^t(\xi_2)}\leq \norminf{\xi_1-\xi_2},\quad \forall t\in\rea,
$$
for any pair of initial conditions $\xi_1,\xi_2\in\Xc$. By replacing ${\phi^t(x)=\varphi(t,x)}$, the proof is complete. 
\end{proof}

\section{ K-topical dynamical systems}\label{sec:dynamicalSystems}

The main result of this section, given in Theorem \ref{th:convergence}, is that for smooth K-topical systems in continuous or discrete-time, each trajectory converges to some stable equilibrium point, if any exists. For the convenience of the reader, we state here this result and postpone its proof at the end of this section, which makes use of several intermediate results.

\begin{thm}[{Convergence}]\label{th:convergence}
~
Consider a K-topical dynamical system on $\Xc\subseteq \rea^n$. If $f$ is $C^1$ and the set of equilibrium point $\Fc$ is not empty, then for any initial condition $\xi\in \Xc$ there exists a stable equilibrium point $x_{\xi}\in\Fc$, such that
\begin{equation*}
    \lim_{t \rightarrow \infty} \varphi(t,\xi)=x_{\xi},\quad \forall \xi\in \Xc.\clt
\end{equation*}
\end{thm}

As the above result is given for a general dynamical system, regardless of which framework is considered (continuous or discrete-time), we need to establish some equivalence relation.
First of all, in Lemma~\ref{lem:glob_ex} we prove that solutions of continuous-time topical systems are unique and exist at all times. Secondly, in Lemma~\ref{lem:eq_discsys} we show how to construct a discrete-time system from a continuous-time topical system with the same asymptotic behavior. Finally, after having proved the stability of each equilibrium point in Lemma~\ref{lem:stability}, the proof of Theorem~\ref{th:convergence} is carried out for the equivalent discretized system.


Let us step back for a moment and focus on the topicality property. Topical systems have been considered for decades in discrete-time,
\begin{equation}\label{eq:disc_sys}
    x(k+1) = f(x(k)),\qquad k\in\nat.
\end{equation}
In this case, the properties of the flow $\varphi$ directly translates into properties of the map $f$ since $\varphi(k,\xi) = f^k(\xi)$ for any initial condition $\xi\in\Xc$ and time $k\in\nat$.
Thus, the asymptotic behavior of the system is studied by considering the iterative behavior of the map $f \equiv \varphi^1$.
%
%

On the other hand, less attention has been paid to continuous-time systems,
 \begin{equation}\label{eq:cont_sys}
\dot{x}(t) = f(x(t)),\qquad t\in\mathbb{R}.
\end{equation}
We show in Lemma~\ref{lem:eq_discsys} that, similarly to the discrete-time case, the asymptotic behavior of the continuous-time system can be inferred from the iterative behavior of its flow $\varphi^T$, for any time discretization ${T\in\rea_{\geq 0}}$ under the assumption of a continuously differentiable vector field. To this aim, we first need to prove in the next lemma that their flow is defined and unique at all times.

\begin{lem}\label{lem:glob_ex}
If a continuous-time system as in eq.~\eqref{eq:cont_sys} is topical and if $f$ is $C^1$, then for any initial condition $\xi\in\Xc$ the flow $\varphi(t,\xi)$ exists for all $t\in\rea_{\geq 0}$ and it is unique.
\end{lem}
\begin{proof}
Since $f\in C^1$, then we have the following facts\footnote{The proof of these standard results can be found in Section 17.2 and Section 17.6 of~\cite{Hirsch12}, respectively.}:
\begin{enumerate}[label=$(\roman*)$]
    \item For any initial condition $\xi\in\Xc$ the flow $\varphi(t,\xi)$ exists in an interval $[0,T]$ and it is unique in it;
    \item the flow $\varphi(t,\xi)$ is $C^1$, i.e., its partial derivatives with respect to time and initial conditions exists and are continuous in the interval of existence $[0,T]$.
\end{enumerate}

By topicality of the system, we can exploit the non-expansiveness property given by Lemma~\ref{lem:non-expansiveness} to ensure that solutions exist for all $t\geq 0$. Consider any initial condition $\xi\in\Xc$ and a subsequent state $\varphi(t^*,\xi_1)$ with $t^*>0$, then we can write
$$
\norm{\varphi(t,\xi) - \varphi(t,\varphi(t^*,\xi)}_{\infty} \leq  \norm{\xi-\varphi(t^*,\xi)}_{\infty},\quad \forall t\in\mathbb{T}.
$$

The above relation says that the flow is Lipschitz and therefore it does not diverge in finite time.
This, jointly with fact $(ii)$ that the flow is continuous and differentiable in the interval of existence, ensures that the existence and uniqueness of the solutions stated in $(i)$ hold in the interval $[0,\infty)$, thus completing the proof.
\end{proof}

\begin{lem}\label{lem:eq_discsys}
Consider a continuous-time system as in eq.~\eqref{eq:cont_sys}. Let $\varphi(t,\xi)$ be the flow of system, fix an arbitrary time step $T>0$, define the map $g(\cdot) = \varphi(T,\cdot)$ and consider the discrete-time dynamical system defined by
\begin{equation}\label{eq:eq_discsys}
y(k+1) = g(y(k)),\qquad \forall T>0.
\end{equation}

If the continuous-time system is topical and if its vector flow $f$ is $C^1$, and if the initial states of the continuous-time and discrete-time systems coincide, i.e., $x(0)=y(0) =\xi$, then
$$
\lim_{t\rightarrow\infty} x(t) = \lim_{k\rightarrow\infty } y(k).
$$
\end{lem}
\begin{proof} Having shown the existence and uniqueness of flows in Lemma~\ref{lem:glob_ex}, the systems satisfies the so-called \emph{group law} (cfr.~\cite[Section 7.1]{Coppel65}),
$$
\varphi(q,\varphi(p,\xi)) = \varphi(p+q,\xi)),\qquad 
$$

By selecting $p=q=T\in\mathbb{R}$ we can write $\varphi(T,\varphi(T,\xi)) = \varphi(2T,\xi)$. Thus, letting $g(\cdot) = \varphi(T,\cdot)$, consider the discrete-time system in eq.~\eqref{eq:eq_discsys}, and write
$$
x(kT) = \varphi(kT,\xi) = g^k(\xi) = y(k),
$$
from which the statement of the theorem derives trivially.
\end{proof}

Due to Lemma~\ref{lem:eq_discsys}, regardless of whether the system under consideration is continuous or discrete in time, one can equivalently study its asymptotic behavior by means of a discrete-time system as in eq.~\eqref{eq:eq_discsys}. This enables us to prove in the next Lemma~\ref{lem:stability} that each equilibrium point of topical systems is stable and, consequently, the main result anticipated at the beginning of this section.

\begin{lem}\label{lem:stability}
~
If a dynamical system is topical and if $f$ is $C^1$, then every equilibrium point $x_e\in \Fc$ is stable.
\end{lem}

\begin{proof}
~
By Lemma~\ref{lem:eq_discsys}, consider the discrete-time system as defined in eq.~\eqref{eq:eq_discsys} for $T=1$.
Let $x_e\in\Fc$ be an equilibrium point, then all points $x_e+c \mathbf{1}$ with $c\in \rea$ are equilibrium points. Therefore, for any neighborhood $\mathcal{W}$ of $x_e$, one can find two equilibrium points belonging to this neighborhood $a,b\in \mathcal{W}\cap \Fc$ such that $a<x_e<b$ and $[a,b]\subset \mathcal{W}$. By monotonicity property, it holds
$$
a=g^k(a)\leq g^k(x) \leq g^k(b)=b,\qquad \forall x\in [a,b],k\in\nat.
$$
The proof is complete by observing $g([a,b])\subset [a,b]$. 
\end{proof}


\begin{proof}{Proof of Theorem~\ref{th:convergence}.}
~
By Lemma~\ref{lem:eq_discsys}, consider the discrete-time system as defined in eq.~\eqref{eq:eq_discsys} for $T=1$.
By assumption, the map $g:\Xc\rightarrow \Xc$ is topical and thus by Lemma~\ref{lem:non-expansiveness} it is non-expansive under the sup-norm. Trajectories of sup-norm non-expansive maps have been proved either to be all unbounded or to converge to a periodic point\footnote{This was first proved by Lemmens in~\cite{Lemmens03} and it can be found in its recent book~\cite[Chapter 4]{LemmensNussbaum2012}.}: the trajectory ${\Tc(\xi)=(g^k(\xi))_{k\in \nat}}$ starting at $\xi \in \Xc$ is said to converge to the periodic point $x_{\xi}$ if
$$
\lim_{k\rightarrow\infty} g^{kp}(\xi) = x_{\xi},
$$
where $p\in \nat$ is the period of $x_{\xi}$. 
Since by assumption there exists at least one equilibrium point ${x_e\in \Fc(g)}$, then for any point $\xi\in\Xc$, there exists a periodic point $x_{\xi}$ which the trajectory through $\xi$ converges~to.

We claim that for topical systems possessing the additional property of type-K monotonicity, all periodic points are equilibrium points. This leads to the conclusion that the system always converges to an equilibrium point that is stable due to Lemma~\ref{lem:stability}, thus completing the proof. In proof of the claim, we make use of the concept of \emph{limit set} of a point $\xi\in\Xc$, which we denote by
$$\Omega(\xi)=\{x_{\xi},g^{1}(x_{\xi}),\cdots,g^{p-1}(x_{\xi})\}.$$

Moreover, we consider the tightest lower bound ${\ell(\xi)=[\ell_1(\xi),\cdots,\ell_{|\xi|}(\xi)]}$ to such limit set, where $|\xi|$ denotes the number of components of $\xi$, whose formal definition is given next
$$
\ell_i(\xi) = \min_{x\in\Omega(\xi)} x_i,\quad \forall i=1,\ldots,|\xi|. 
$$
With these definitions, the claim \virg{all periodic points are equilibrium points} can be equivalently stated as follows
\begin{equation}\label{eq:claim}
    \abs{\Omega(\xi)}=1,\qquad \forall \xi \in \Xc,
\end{equation}
where $\abs{\cdot}$ denotes the cardinality of a set.
By definition of $\ell(\xi)$, for any $i=1,\ldots,|\xi|$ there exists ${y\in \Omega(\xi)}$ such that ${\ell_i(\xi) = y_i}$.
Thus, for each $x\in \Omega(\xi)$ either $\ell_i(\xi)<x_i$ or $\ell_i(\xi)=x_i$.
Assume that $\ell_i(\xi)<x_i$. Since $f$ is type-K monotone, it holds  ${\ell_i(\xi) < f^k_i(x)}$ for all $k\geq 0$. 
Since $ x,y\in \Omega(\xi)$, then there exists a $k$ such that ${f^k(x)=y}$, it follows that $\ell_i(\xi)<y_i$, which is a contradiction. 
Thus, for all $x\in\Omega(\xi)$ it holds that $\ell_i(\xi)=x_i=y_i$, i.e., all points in the limit set $\Omega(\xi)$ have the same $i$-th component.
The same reasoning holds for all components $i=1,\ldots,|\xi|$, thus proving the claim in eq. \eqref{eq:claim} and completing the proof.
\end{proof}

\section{How to verify K-topicality}

\subsection{K-topicality in continuous-time}
Angeli and Sontag have proved that the plus-homogeneity of a continuous-time system can be verified by only looking at the function $f$, as remarked next.

\begin{rem}\label{rem:phomo_cont}
\cite[Lemma 3.1]{Angeli06} A continuous-time system as in eq.~\eqref{eq:cont_sys} on $\Xc\in\rea^n$ is plus-homogeneous if
\begin{equation*}
    f(\xi+\alpha\bone) = f(\xi),\qquad \forall \alpha \in \mathbb{R},\forall \xi\in\Xc.
\end{equation*}
\end{rem}

On the other hand, verifying the monotonicity of a system is not an easy task. For continuous-time dynamical systems whose vector field is continuously differentiable, a necessary and sufficient condition to ensure monotonicity is the well-known \emph{Kamke condition}, which dates back to the 30s and the work of Kamke in~\cite{Kamke32}, see~Proposition 1.1 in~\cite{Smith08}\footnote{Note that in standard books, such as those of Smith~\cite{Smith08} and Coppel \cite{Coppel65}, \virg{monotonicity} is referred to as \virg{type-K}, even if this notation has been lost in the current literature. In this paper, we recover the notation \virg{type-K} with a different meaning.}.

\begin{thm}[Kamke condition]\label{thm:kamke_cond}
A continuous-time system as in eq.~\eqref{eq:cont_sys} on $\Xc\in\rea^n$ is monotone if and only if for any two points $a,b \in \Xc$ such that $a\leq b$ the following holds
$$
\forall i: \: a_i=b_i \quad \Rightarrow \quad  f_i(a)\leq f_i(b).
$$

\end{thm}

It should be noted that from the previous theorem it follows that any scalar continuous-time system is monotone, since the condition is satisfied trivially for $n=1$.

\begin{rem}\label{rem:kamke_cond_sca}
Any scalar continuous-time system as in eq.~\eqref{eq:cont_sys} on $\Xc\in\rea$ is monotone.
\end{rem}

For continuous-time systems with a continuously differentiable vector field, the Kamke condition turns out to be equivalent to a specific sign structure of the Jacobian matrix, see~Remark 1.1 in~\cite{Smith08} for a simple proof.

\begin{cor}\label{cor:mon_jac}
Consider a continuous-time system~\eqref{eq:cont_sys} on $\Xc\subseteq \rea^n$ whose vector field $f$ is $C^1$. The system is monotone if and only if Jacobian matrix is Metzler, i.e.,
$$
\frac{\partial f_i(x)}{\partial x_j}\geq 0,\qquad  i\neq j,\quad x\in\Xc
$$
\end{cor}

In the following, we show that for a continuous-time system whose vector field is continuously differentiable, monotonicity is equivalent to type-K monotonicity, thus proving that the same sign structure of the Jacobian matrix is also a necessary and sufficient condition for type-K monotonicity. 

\begin{thm}\label{th:monotoneK}
Consider a continuous-time system in $\Xc\subseteq \rea^n$ with dynamics
\begin{equation}\label{eq:sys}
\dot{x}(t) = f(x(t)),\qquad t\in \mathbb{R}_{\geq 0}.
\end{equation}
If the system is monotone and its vector field $f$ is $C^1$, then the system is type-K monotone.
\end{thm}

\begin{proof}
Consider a non-negative vector $v\in\mathbb{R}^n_{\geq 0}$ and denote by $x(t)$ and $z(t)$ the solutions of the monotone system in eq.~\eqref{eq:sys} at time $t\in\mathbb{R}$ with initial conditions $x(0)\in\rea^n$ and $z(0)=x(0)+v$, respectively, i.e.,
\begin{equation*}
   x(t) = \varphi(t,x(0)),\qquad z(t) = \varphi(t,x(0)+v), 
\end{equation*}
where $\varphi$ is the flow of the monotone system in eq. \eqref{eq:sys}. 
Without loss of generality, assume that both solutions $x(t)$ and $z(t)$ exists in an interval $[0,T^*]$ with $T^*\in\rea_{> 0}$.

The monotonicity of the system implies that the order between the initial conditions, $x(0)\leq z(0)$, must be preserved by the solutions at all times, i.e.,
\begin{equation}\label{eq:mono_proof}
x(t)\leq z(t), \qquad t\in[0,T^*].
\end{equation}

To prove the type-K monotonicity of the system, we need to show that if there is a strict order in the $i$-th component, i.e., ${x_i(0)<z_i(0)}$, which is equivalent to ${v_i>0}$, then such order is preserved at all times, i.e., for ${t\in[0,T^*]}$ it holds
\begin{equation}\label{eq:typek_proof}
v_i>0 \quad \Rightarrow \quad x_i(t)<z_i(t).
\end{equation}

At $t=0$ eq.~\eqref{eq:typek_proof} holds because
${x_i(0)<x_i(0)+v_i=z_i(0)}$.
Now, since $f$ is $C^1$, then both solutions $x(t)$ and $z(t)$ are also $C^1$, and thus there exists an interval of time $[0,t^*)$ of positive measure, i.e.,  $0<t^*\leq T^*$, in which eq.~\eqref{eq:typek_proof} holds.

Finally, we aim to prove that eq.~\eqref{eq:typek_proof} holds for all $t\in [0,T^*]$ by contradicting the following
\begin{equation}\label{eq:absurd}
\exists \:T\in [t^*,T^*]: \quad x_i(T)=z_i(T).
\end{equation}

Denoting $a_{-i}\in\rea^{n-1}$ the vector of $(n-1)$ elements obtained from vector $a\in\mathbb{R}^n$ by removing the $i$-th component, i.e., $a_{-i} = [a_1,\ldots,a_{i-1},a_{i+1},\ldots,a_n]^\intercal $, we can say that the $i$-th component of $x(t)$ is solution of the differential equation
\begin{equation}\label{eq:diff_eq}
\dot{x}_i(t) = f_i(x_i(t),x_{-i}(t)).
\end{equation}
where $x_{-i}(t)$ is treated as an exogenous input.
Similarly, the $i$-th component of $z(t)$ is solution of
$$
\dot{z}_i(t)=f_i(z_i(t),z_{-i}(t)).
$$

Moreover, from the monotonicity of the system in eq. \eqref{eq:mono_proof}, which implies $z_{-i}(t)\geq x_{-i}(t)$, and from Corollary~\ref{cor:mon_jac}, which states that the map $f_i$ is a nondecreasing function in all variables other than the $i$-th, i.e., ${f_i(z_i(t),z_{-i}(t))\geq f_i(z_i(t),x_{-i}(t))}$, it follows that
$z_i(t)$ is also a solution of the differential inequality
\begin{equation}\label{eq:new_ineq}
\dot{z}_i(t) \geq f_i(z_i(t),x_{-i}(t)).
\end{equation}

We now operate a time reversal and a time shif by letting $\tau = T-t$. We denote $x_i^{rev}(\tau)=x_i(T-\tau)$ and ${z_i^{rev}(\tau)=z_i(T-\tau)}$ the reversed solutions. 
By this change of variables we compute
\begin{align*}
\dot{x}_i^{rev}(\tau) = \frac{d x_i^{rev}(\tau)}{d \tau } = \frac{d x_i(T-\tau)}{d \tau }= - \dot{x}_i(T-\tau),
\end{align*}
and from eq.~\eqref{eq:diff_eq} we derive that
$x_i^{rev}(\tau)$ is solution of
\begin{equation}\label{eq:diff_eq2}
\dot{x}_i^{rev}(\tau) = - f_i(x_i^{rev}(\tau),x_{-i}^{rev}(\tau)).
\end{equation}

With similar steps, from eq. \eqref{eq:new_ineq} we derive that
$z_i^{rev}(\tau)$ is a solution of
\begin{equation}\label{eq:diff_ineq2}
\dot{z}_i^{rev}(\tau) \leq - f_i(z_i^{rev}(\tau),x_{-i}^{rev}(\tau)).
\end{equation}

Assuming that eq.~\eqref{eq:absurd} holds, at $\tau = 0$ the two solutions are equal, namely $x_i^{rev}(0)=z_i^{rev}(0)$, in fact
$$
x_i^{rev}(0) = x_i(T) = z_i(T) = z_i^{rev}(0),
$$
and since $z_i^{rev}(\tau)$ is a solution of the differential inequality~\eqref{eq:diff_ineq2}, while $x_i^{rev}(\tau)$ is solution of~\eqref{eq:diff_eq2}, then
\begin{equation}\label{eq:rev_contradicttion}
{z_i^{rev}(\tau) \leq x_i^{rev}(\tau)},\qquad \tau \in [0,T^*],
\end{equation}
which, at $\tau = T$, leads to
$$
z_i(0) = z_i^{rev}(T) \leq x_i^{rev}(T) = x_i(0).
$$

This leads to a contradiction since $v_i>0$ by eq. \eqref{eq:typek_proof} and therefore $z_i(0)=x_i(0)+v_i>x_i(0)$. Therefore, eq.~\eqref{eq:absurd} does not hold, and eq.~\eqref{eq:typek_proof} holds instead for all $t\in [0,T^*]$, completing the proof of the theorem.
\end{proof}

\begin{rem}It is important to remark that type-K monotonicity always implies monotonicity, but not vice versa. In particular, Theorem~\ref{th:monotoneK} leads to the following statements:
\begin{itemize}
    \item $f$ is type-K monotone $ \Rightarrow$ $f$ is monotone;
    \item $f$ is monotone and $f\in C^1$ $\Rightarrow$ $f$ is type-K monotone;
    \item $f$ is monotone and $f\notin C^1$ $\not\Rightarrow$ $f$ is type-K monotone.
\end{itemize}
\end{rem}

The above remark emphasizes the role of Theorem~\ref{th:monotoneK}, which states that if a monotone continuous-time system has a continuously differentiable vector field, then it is type-K monotone.
In other words, under the assumption of a continuously differentiable vector field, all monotone systems are also type-K monotone. Consequently, all results provided in this paper for type-K monotone systems apply to smooth monotone systems.

As a corollary to Theorem~\ref{thm:kamke_cond}, we restate Corollary~\ref{cor:mon_jac} in the particular case of type-K monotone systems with continuously differentiable vector fields.
\begin{cor}\label{cor:kamkeKcond}
~
Consider a continuous-time system~\eqref{eq:cont_sys} on $\Xc\subseteq \rea^n$ whose vector field $f$ is $C^1$. The system is type-K monotone if and only if Jacobian matrix is Metzler, i.e.,
$$
\frac{\partial f_i(x)}{\partial x_j}\geq 0,\qquad  i\neq j,\quad x\in\Xc
$$
\end{cor}

\subsection{K-topicality in discrete-time}

%
Verifying the plus-homogeneity of a discrete-time system by only looking at the function $f$ can be done by directly applying Definition~\ref{def:p-homo}, as remarked next.

\begin{rem}\label{rem:phomo_disc}
A discrete-time system as in eq.~\eqref{eq:disc_sys} on ${\Xc\subseteq \rea^n}$
is plus-homogeneous if and only if 
\begin{equation*}\label{eq:phomo_dt}
f(\xi+\alpha\bone) = f(\xi)+\alpha\bone,\qquad \forall \alpha \in \mathbb{R},\forall \xi\in\Xc.
\end{equation*}
\end{rem}

As a counterpart to the Kamke condition given in Theorem~\ref{thm:kamke_cond}, we provide a necessary and sufficient condition for type-K monotonicity in the case of discrete-time systems, which we denote \emph{Kamke-like condition}.

\begin{thm}[Kamke-like condition]\label{thm:kamke_like_cond}
A discrete-time system as in eq.~\eqref{eq:cont_sys} on $\Xc\in\rea^n$ is monotone if and only if for any two points $a,b \in \Xc$ the following holds
\begin{equation}\label{eq:cond1}
a\leq b \Rightarrow f_i(a)\leq f_i(b),
\end{equation}
and it is type-K monotone if and only if it further satisfies
\begin{equation}\label{eq:cond2}
\forall i: \quad a_i<b_i \Rightarrow f_i(a)< f_i(b).
\end{equation}
\end{thm}

\begin{proof}
The solution of a discrete-time system at time $k\in\nat$ is equal to the $k$-th composition of the map $f$, i.e., ${\varphi(k,\xi)=f^k(\xi)}$ for any $\xi\in\Xc$. With this notion, the proof is a straightforward application of Definition~\ref{def: K-topicality}, where ${\varphi(k,\xi)=f^k(\xi)}$.
\end{proof}

For discrete-time systems with a continuously differentiable vector field, the Kamke-like condition turns out to be equivalent to a specific sign structure of the Jacobian matrix, similar to what happens in continuous-time.
A preliminary sufficient condition was given in~\cite[Proposition 9]{Deplano20}, while next, we provide a necessary and sufficient condition.
\begin{thm}\label{thm:kamkelikeKcond}
~
Consider a discrete-time system as in eq.~\eqref{eq:disc_sys} on $\Xc\subseteq \rea^n$ whose map $f$ is $C^1$. The system is type-K monotone if and only if the Jacobian matrix is non-negative,~i.e., 
$$
\frac{\partial f_i(x)}{\partial x_j}\geq 0,\qquad x\in\Xc,
$$
with a strictly positive diagonal almost everywhere, i.e.,
$$
\frac{\partial f_i(x)}{\partial x_i}> 0\quad x\in \Xc \setminus S,
$$
where $S$ is a set of measure zero.
\end{thm}
\begin{proof}
Consider two vectors $y,z\in \Xc$ such that ${y\leq z}$ and, without lack of generality, let $v\in\mathbb{R}^n_{\geq0}$ be such that ${z=y+v}$. For any $i=1,\ldots,n$, we can compute the directional derivative of $f_i$ at $y$ along $v$ by means of the limit definition,
\begin{equation}\label{eq:limit}
\nabla _v f_i(y) = \lim_{\varepsilon\rightarrow 0} \frac{f_i(y+\varepsilon v) - f_i(y)} {\varepsilon}.
\end{equation}

It follows that eq.~\eqref{eq:cond1} is equivalent to a non-negative directional derivative $\partial f_i(y)/\partial v \geq 0$, in fact ${y\leq y+\varepsilon v \Rightarrow f_i(y) \leq f_i(y+\varepsilon v)}$. Since $f\in C^1$, we can write the directional derivative in terms of partial derivatives, 
\begin{equation}\label{eq:gadient}
\nabla _v f_i(y) =\sum_{j=1}^n \frac{\partial f_i(y)}{\partial x_j}\cdot \frac{v_j}{|v|}\geq 0,
\end{equation}
and noticing that the above relation must hold for any ${v\in \mathbb{R}^n_{\geq 0}}$, one infers that the non-negativeness of the Jacobian matrix is necessary and sufficient for eq.~\eqref{eq:cond1}. 
Now, eq.~\eqref{eq:cond2} is equivalent to the fact that function $f_i$ is a strictly increasing function with respect to the variable $x_i$, which in turn is equivalent to the requirement that the partial derivative of $f_i$ with respect to $x_i$ can be zero at most in a set $S$ of measure zero in $\Xc$, cfr.~\cite[Section I.1]{Szarski65}.
This completes the proof.
\end{proof}

\section{K-topical multi-agent systems}\label{sec:multiAgent}

In this section, we exploit the result presented in the previous sections to study the stability of nonlinear K-topical Multi-Agents Systems (MASs). As a special case, we provide sufficient conditions enabling the agents to achieve consensus asymptotically.

We consider MASs composed of $n\in\nat$ agents modeled as dynamical systems with scalar state $x_i(t)\in\mathbb{R}$ whose pattern of interaction is described by a directed graph ${\mathcal{G}=(\mathcal{V},\mathcal{E})}$, and evolving either in discrete-time
\begin{equation}
\label{eq:locdiscretesystem}
    x_i(k+1) = f_i\left(x_i(k),x_j(k):\Nc_i\right), \qquad k\in \nat.
\end{equation}
or in continuous-time
\begin{equation}
\label{eq:loccontinuoussystem}
    \dot{x}_i(t)  = f_i\left(x_i(t),x_j(t):\Nc_i\right), \qquad t\in \rea,\\
\end{equation}
%
%

We first state our main result for discrete-time MASs. This result provides necessary and sufficient conditions on the local interaction rule $f_i$ of the single agent, which can be different from the others, ensuring that the overall MAS results to be a K-topical dynamical system, thus proving its convergence to the equilibrium point set $\Fc\neq \emptyset$ by means of Theorem~\ref{th:convergence}. Moreover, we provide some extra sufficient conditions ensuring that the equilibrium point set $\Fc$ coincides with the consensus set
\begin{equation}\label{eq:consensus_space}
\mathcal{C}=\{\alpha\bone: \alpha\in \rea \},
\end{equation}
thus solving the consensus problem for nonlinear K-topical MASs. The proposed sufficient condition is graph theoretical and based on the graph $\Gc$ describing the pattern of interconnections among the agents: it requires that there exists a globally reachable node in $\Gc$ and that consensus states are equilibrium points.
We collect in the next lemma some useful intermediate results which are instrumental to the proof of our main results. 

\begin{lem}\label{lem:extracond}
Consider a K-topical discrete-time system as in eq.~\eqref{eq:disc_sys} on $\Xc\subseteq \rea^n$ whose map $f$ is $C^1$.
%
%
%
Then:

\begin{enumerate}[label=$(\alph*)$]
    \item The Jacobian matrix $J_f=\{\partial f_i/ \partial x_j\}$ is stochastic everywhere, i.e.,
    $$
    J_f(\xi)\bone = \mathbf{1},\qquad \forall \xi\in \Xc;
    $$
    \item The set of equilibrium points $\mathcal{F}$ is either empty or closed and convex, i.e., for all $x,y \in \mathcal{F}$ then ${\alpha x + (1-\alpha y)\in\mathcal{F}}$ for all $\alpha \in [0,1]$;
    \item If $f(\mathbf{0})=\mathbf{0}$ then all consensus points are equilibrium points, i.e., $\mathcal{C} \subseteq \mathcal{F}$.
    \end{enumerate}
\end{lem}

\begin{proof}
Statement $(a)$ is proved by exploiting Remark~\ref{rem:phomo_disc} to the definition directional derivative at $\xi\in\Xc$ along the vector $\bone\in\mathbb{R}^n$, as follows,
\begin{align*}
J_f(\xi)\bone &=
\lim_{h\rightarrow 0} \frac{f(\xi+h\bone)-f(\xi)}{h}\\
&=\lim_{h\rightarrow 0} \frac{f(\xi)+h\bone-f(\xi)}{h}
=\lim_{h\rightarrow 0} \frac{h\bone}{h}=\bone.
\end{align*}
Statement $(b)$ is a straightforward application of \cite[Theorem 1]{Ferreira96} in the euclidean normed space $(\Xc,\norminf{\cdot}$).

Statement $(c)$ is a consequence of the plus-homogeneity property, in fact
\begin{align*}
f(\bzero+\alpha\bone) &= f(\bzero) + \alpha\bone,& \forall \alpha \in \rea ,\\
&\Downarrow &
\\
f(\alpha\bone) &= \alpha\bone, & \forall \alpha \in \rea .
\end{align*}
\end{proof}

\begin{thm}[{Discrete-time MAS}]\label{th:consensus_DT}
~
Consider a discrete-time MAS as in eq.~\eqref{eq:locdiscretesystem} on $\Xc\subseteq \rea^n$ whose map is continuously differentiable. If the set of local interaction rules ${f_i:\Xc\rightarrow\rea }$, with ${i= 1,\ldots,n}$, satisfies the next conditions:
\begin{enumerate}[label=$(\roman*)$]
\item $\partial f_{i}/\partial x_{i}>0$ and $\partial f_{i}/\partial x_{j}\geq 0$ a.e. for $i\neq j$;
\item $f_i(x+\alpha) = f_i(x) + \alpha$ for any $\alpha\in\rea $;
\end{enumerate}
then the MAS converges asymptotically to one of its equilibrium points, if any, for any initial state $x(0)\in\Xc$.

If it further satisfies
\begin{enumerate}[label=$(\roman*)$]
\setcounter{enumi}{2}
\item $f_i(\bzero)=0$;
\item The graph $\Gc$ has a globally reachable node;
\end{enumerate}
then the MAS converges asymptotically to a consensus state for any initial state $x(0)\in\Xc$.
\end{thm}

\begin{proof}
The MAS is a K-topical system: condition $(i)$ implies type-K monotonicity by Theorem~\ref{thm:kamkelikeKcond} and condition $(ii)$ implies plus-homogeneity, as underlined in Remark~\ref{rem:phomo_disc}.
If the system has at least one equilibrium point, we can exploit the result in Theorem~\ref{th:convergence} to establish that for any initial conditions $x(0)\in\Xc$, the state trajectories of the MAS converge to one of its equilibrium points in $\Fc$, completing the first part of the proof.

If condition $(iii)$ implies $\mathcal{C}\subseteq \mathcal{F}$ by Lemma~\ref{lem:extracond}, we are going to prove that condition $(iv)$ further implies that $\mathcal{C} \equiv \mathcal{F}$ .
The graph $\Gc$ is aperiodic due to condition $(i)$ which ensures the presence of a self-loop at each node and it contains a globally reachable node due to condition $(iv)$.
Since the Jacobian matrix $J_f$ is row-stochastic everywhere by Lemma~\ref{lem:extracond}, we can exploit the widely known Theorem 5.1 in~\cite{Bullo18} and conclude that $J_f$ has a simple unitary eigenvalue with corresponding eigenvector equal to $\bone$, unique up to a scaling factor.
Now, assume there exist an equilibrium point $x_e \neq \beta\bone$. By Lemma~\ref{lem:extracond}, all points $\alpha x_e + (1-\alpha)c\bone$ with $\alpha\in[0,1]$ and $c\in\mathbb{R}$ are also equilibrium points due to the convexity of the set of equilibrium points. Therefore, all consensus points $c\bone$ with $c\in\mathbb{R}$ have two eigenvectors, the vector of ones $\bone$ and the vector $x_e$.
In other words, all consensus points have a unitary eigenvalue with multiplicity strictly greater than one, which is in contradiction with \cite[Theorem 5.1]{Bullo18}.
Therefore, there does not exist any point $x_e\in\mathcal{F}\setminus \Xc$, completing the second part of the proof.

\end{proof}

In the next theorem, we exploit Lemma~\ref{lem:eq_discsys} to generalize the previous result to continuous-time MASs.

\begin{thm}[{Continuous-time MAS}]\label{th:consensus_CT}
~
Consider a continuous-time MAS~\eqref{eq:loccontinuoussystem} on $\Xc\subseteq \rea^n$ whose vector field is continuously differentiable. If the set of local interaction rules ${f_i:\Xc\rightarrow\rea }$, with ${i= 1,\ldots,n}$, satisfies the next conditions:
\begin{enumerate}[label=$(\roman*)$]
\item $\partial f_{i}/\partial x_{j}\geq 0$ for $i\neq j$;
\item $f_i(x+\alpha) = f_i(x)$ for any $\alpha\in\rea$;
\end{enumerate}
then the MAS converges asymptotically to one of its equilibrium points, if any, for any initial state $x(0)\in\Xc$.

If it further satisfies
\begin{enumerate}[label=$(\roman*)$]
\setcounter{enumi}{2}
\item $f_i(\bzero)=0$;
\item The graph $\Gc$ has a globally reachable node;
\end{enumerate}
then the MAS converges asymptotically to a consensus state for any initial state $x(0)\in\Xc$.
\end{thm}

\begin{proof}
The MAS is K-topical: condition $(i)$ implies type-K monotonicity by K Corollary~\ref{cor:kamkeKcond} and condition $(ii)$ implies plus-homogeneity, as underlined in Remark~\ref{rem:phomo_cont}.
By means of Lemma~\ref{lem:eq_discsys}, we can study its asymptotic behavior by studying the following discrete-time system
$$
x(k+1) =g(x(k)) = \varphi(1,x(k)),\qquad k\in\nat.
$$

The proof is complete by noticing that all conditions $(i)-(iv)$ of Theorem~\ref{th:consensus_DT} hold.
\end{proof}

\section{Examples of application}\label{sec:examples}

\subsection{The multiplicative framework}\label{sec:homofield}
Topical systems are closely related to monotone homogeneous vector fields on the (strictly) positive orthant $\rea_{> 0}^n$. The whole space $\rea^n$ can be put in bijective correspondence with $\rea_{\geq 0}^n$ via the mutually inverse bijection ${\exp:\rea^n\rightarrow\rea_{> 0}^n}$ and ${\log:\rea_{> 0}^n\rightarrow\rea^n}$, which are to be intended as component-wise operations (cfr. \cite[Section 2.7]{LemmensNussbaum2012}).
If ${f:\rea \rightarrow \rea }$ is any self-map of the real vector space, let ${g:\rea_{> 0}^n\rightarrow \rea_{> 0}^n}$ denote the function ${g=\exp \circ f \circ \log}$. 
The properties of $\exp$ and $\log$ show that monotonicity and plus-homogeneity of a system ruled by function~$f$ correspond to the following properties of a system ruled by function~$g$,
\begin{align}\label{eq:mono_p}
 \xi_1 \leq \xi_2 &\Rightarrow \varphi(t,\xi_1)\leq \varphi(t,\xi_2),\quad \forall \xi_1,\xi_2 \in \rea_{> 0}^n, \\
 \varphi(t, \alpha\xi) &= \alpha\varphi(t,\xi) ,\quad \forall \xi\in \rea^n,\forall \alpha\in\rea_{\geq 0} \label{eq:homo_p},
\end{align}
%

While the monotonicity property in eq.~\eqref{eq:mono_p} (as well as type-K monotonicity) is naturally inherited, the plus-homogeneous property corresponds to the homogeneity property in eq.~\eqref{eq:homo_p}.
%
Consequently, the results provided in this paper for K-topical systems in $\rea^n$ have equivalent multiplicative formulations, i.e., they apply to type-K monotone and homogeneous systems in ${\Xc \subseteq \rea^n_{\geq 0}}$, both in discrete-time~\cite{Deplano20} and continuous-time. We leave it to the reader to formulate any dual~result.

\subsection{Dynamical systems}

\textbf{Chemical reactions.} The most studied case of K-topical systems in continuous-time is that of well-mixed and isothermal chemical reactions \cite{Angeli08, Hu10}. 

Let $s(t)\in\mathbb{R}^n$ denote the vector specifying the concentrations of $m$ chemical species, $\Gamma\in\mathbb{R}^n\rightarrow\rea^m$ be the stoichiometry matrix and $h:\rea^m_{\geq 0}\rightarrow \rea^n$ be a function which provides the vector of reaction rates for any given vector of concentrations, then the dynamics of the system is given by
$$
\dot{s}(t) = \Gamma h(s(t)).
$$

Using the reaction coordinates $x(t)$ such that ${s(t) = \Gamma x(t)}$, the system dynamics becomes $
\Gamma\dot{x}(t) = \Gamma h(\Gamma x(t))$, and one can infer the stability of this system by studying the system $\dot{x}(t) = h(\Gamma x(t)),$
which is K-topical by Theorem~\ref{th:monotoneK}: it is monotone and plus-homogeneous, with continuously differentiable vector~field.


%

\textbf{Max-plus maps.} Important examples of topical systems are those ruled by max-plus maps.
To introduce these maps let $\rea_{\infty}=\rea\cup\{-\infty\}$ denote the \emph{max-plus semi-ring} and let $A=\{a_{ij}\}$ be a $n\times n$ matrix with entries from $\rea_{\infty}$ and suppose that for each $i$ there exists $j$ such that $a_{ij}\neq -\infty$.
A max-plus map $f:\rea^n\rightarrow\rea^n$ is defined by
$$
f_i(\xi)=\max_j\{a_{ij}+x_j\},\quad \forall x\in\rea^n,i=1,\ldots,n.
$$

It is easy to verify that discrete-time systems $x(k+1)=f(x(k))$ ruled by a max-plus map are K-topical. 
%
Applications of max-plus maps arise in several fields, such as optimal control~\cite{Akian94},
decentralized estimation~\cite{Deplano21}, discrete event systems~\cite{Gunawardena2003}, and many others.

\textbf{Stochastic games. }
Another remarkable example are stochastic games, which are two-player zero-sum games \cite{Grammatico17,Belgioioso18} and go back to Shapley~\cite{Shapley53, Sorin03}.
The dynamic programming
method for computing the value of a stochastic game also leads to
a topical map.

Consider a two-player zero-sum game with finite state space $\mathcal{S}=\{1,\ldots,n\}$ and, for every state $i\in \mathcal{S}$, action spaces $\mathcal{A}^1_i$ for player $1$ and $\mathcal{A}^2_i$ for player $2$, transition probabilities $p(j|i,a_1,a_2)$ and transition payments $r(i,a_1,a_2)$ with ${i,j\in\mathcal{S}}$, $a_1\in\mathcal{A}^1_i$ and $a_2\in\mathcal{A}^2_i$.

Its study involves the Shapley operator given by
$$
f_i(x)=\min_{a_1\in\mathcal{A}^1_i}\max_{a_2\in\mathcal{A}^2_i}\{r(i,a_1,a_2)+\sum_{j\in\mathcal{S}}p(j|i,a_1,a_2)x_j\},
$$
where $x_i\in\mathbb{R}$ denotes the final reward at state $i\in\mathcal{S}$.
The reward at $x(k)$ of the game after $k$ steps is determined recursively, using dynamical programming principle ${x(k+1)=f(x(k))}$. It can be noticed that the Shapley operator is both monotone and plus-homogeneous, and thus is topical. Moreover, if the probability $p(i|i,a_1,a_2)$ is positive, which is a natural assumption, then the system is also type-K monotone and thus K-topical. 

\begin{rem}
A smooth version of the max-function used in the previous examples, which does not affect the K-topicality, can be obtained through the approximation shown next, usually called \emph{softmax}~\cite{Gao17},
$$
\alpha\text{-}\max(x) = \frac{\sum_{i=1}^n x_ie^{\alpha x_i}}{\sum_{i=1}^n e^{\alpha x_i}},\quad \alpha > 0,
$$
for which $ \lim_{\alpha\rightarrow\infty } \alpha\text{-}\max(x) = \max (x) $.
\end{rem}


\subsection{Multi-agent systems}

The most common consensus algorithms for discrete and continuous-time single-integrator multi-agent systems 
\begin{equation}\label{eq:sing_int}
\dot{x}_i(t) = u_i,\qquad x_i(k+1) = x_i(k) + \varepsilon_i u_i(t).
\end{equation}
with $\varepsilon>0$ are given by the following control input
\begin{equation}\label{eq:lin_prot}
u_i = \sum_{j\in\Nc_i}\left(x_j-x_i\right).
\end{equation}

It is easy to verify that the standard consensus protocol makes the system K-topical. 
Many variations of eq.~\eqref{eq:lin_prot} have been proposed in several applications, such as formation control in multi-vehicle systems~\cite{Fax04}, the modeling of the emergent flocking behavior~\cite{Vicsek12}, optimization algorithms~\cite{Nedic10}, and many others.
It is remarkable that K-topicality is preserved if one considers nonlinearities of the following type\cite{Munz11}\footnote{Similar results hold also if the nonlinearity is applied after the summation is operated, {\scriptsize ${u_i = h_i\left(\sum_{j\in\Nc_i}\left(x_j-x_i\right)\right)}$}, \cite{Xu13}.}, 
\begin{equation}\label{eq:nonlin_prot}
u_i = \sum_{j\in\Nc_i}h_{ij}\left(x_j-x_i\right),
\end{equation}
under some mild conditions discussed next. We point out that the generality of our approach allows the local interaction rule of the agents to be possibly different from the others, thus enabling the study of heterogeneous multi-agent systems, which is still today a topic of great interest in our community~\cite{Zuo18, Du20, Zhao21}.

(\emph{Continuous-time})
It has been proved that the system in eq.~\eqref{eq:sing_int} with the linear protocol in eq. \eqref{eq:lin_prot} converges to a consensus state if the graph $\mathcal{G}$ possesses a globally reachable node~\cite[Theorem 7.4]{Bullo18}. By means of Theorem~\ref{th:consensus_CT} we directly generalize this result by considering the nonlinear protocol in eq. \eqref{eq:nonlin_prot} couplings $h_{ij}:\mathbb{R}\rightarrow\mathbb{R}$ satisfying
\begin{itemize}
    \item $\displaystyle \frac{\partial}{\partial x_j} h_{ij}\geq 0$ for all $j\neq i$ and $i\in\mathcal{V}$;
    \item $h_{ij}(0)=0$ for all $i,j\in\mathcal{V}$.
\end{itemize}

A similar result is given in~\cite{ZhiyunLin2007}, where in addition the vector field of the global system is required to meet an extra strict sub-tangentiality condition. It is clear that if the maps are taken as the identity map $h_{ij}(x)=x$, then protocol reduces to the linear one in eq. \eqref{eq:lin_prot}.

(\emph{Discrete-time})
The convergence properties of the system in eq.~\eqref{eq:sing_int} with the linear protocol in eq. \eqref{eq:lin_prot} depends on the parameter $\varepsilon$ and the topological structure of $\mathcal{G}$~\cite[Theorem 5.1]{Bullo18}. In particular, the system reaches consensus if the graph possesses a globally reachable node belonging to an aperiodic component, and if $ \varepsilon_i<\abs{\mathcal{N}_i^{-1}}$.
The condition on $\varepsilon$ ensures that the state transition matrix is row-stochastic and nonnegative. In a similar way, one can find a condition on $\varepsilon$ ensuring that the map $f$ given the nonlinear protocol in eq. \eqref{eq:nonlin_prot} is plus-homogeneous and type-K monotone, given by

$$
\displaystyle \varepsilon_i<\left[\abs{\mathcal{N}_i}\frac{\partial}{\partial x_i} h_{ij}\right]^{-1}.
$$
Such property, jointly with the two presented in the previous paragraph, allows to exploit Theorem~\ref{th:consensus_DT} and prove convergence to a consensus state of the system. 


\textbf{Bounded control inputs. }
As the first example of application, consider the case wherein the control inputs are constrained by a saturating effect~\cite{Yang14, Xie19, Ong21}.
%
%
The problem of designing proper saturating functions $h_{ij}$ such that the consensus protocols are yet qualifiable can be solved by the use of the following function
$$
h_{ij}(x) = s_i\left(\frac{1-e^{-m_ix}}{1+e^{-m_ix}}\right),\quad \forall j\in\mathcal{N}_i
$$
with $s_i,m_i> 0$, which is easily proved to be K-topical\footnote{Note that for the discrete-time case it is further required that $ \varepsilon_i<\left[0.5\cdot m_i\cdot s_i\abs{\mathcal{N}_i}\right]^{-1}$.}.

Notably, the proposed function encompasses several well-known saturating functions:
\begin{itemize}
 \item $h_{ij}(x) = \text{tanh}(x)$ if $s_i=1$ and $m=2$;
 \item $h_{ij}(x) = \text{sign}(x)$ if $s_i=1$ and $m\rightarrow \infty$;
\end{itemize}
Theorems~\ref{th:consensus_DT}-\ref{th:consensus_CT} ensures that a multi-agent system wherein the agents are subject to the above described saturated control action achieves consensus if the underlying graph contains a globally reachable node. 

\textbf{Kuramoto Oscillators. }
The emergence of synchronization or desynchronization in networks of coupled oscillators is another interesting example~\cite{Dorfler14, Deplano20distributed}.

Here, we consider a network of oscillators with the same natural frequency whose angular velocities $x_i(t)$ coupled through their phase differences according to a graph $\Gc$ and coupling functions $h_{ij}$. Weakly-coupled identical limit-cycle oscillators can be well approximated by this canonical model through a phase reduction and averaging analysis, with appropriate coupling functions $h_{ij}$ that are closely related to the phase response curve of the oscillators. Since the phase response curve is a function computed on the periodic limit cycle, it is $2\pi$-periodic and so are the coupling functions $h_{ij}$.

Theorem~\ref{th:consensus_CT} constitutes a new analysis tool for studying synchronization in such networks, where the couplings can be directed and heterogeneous, while they must met the next condition,
\begin{equation}\label{eq:piecewise_mono}
\frac{d}{d\theta}h_{ij}(\theta) =\begin{cases}
>0 & \theta \in (-\alpha,\alpha)\\
<0 & \theta \in (-\pi,-\alpha)\cup(\alpha,\pi)
\end{cases},
\end{equation}
with $\alpha\in[0,\pi]$ and $h_{ij}(0)=0$.
It is easy to notice that letting $a,b$ be any real numbers such that $0\leq b-a\leq \alpha$, then Theorem~\ref{th:consensus_CT} holds for $\mathcal{X}=[a,b]^n\subset \mathbb{R}^n$. In fact, $\mathcal{X}$ is an invariant space wherein all conditions of the theorem are satisfied if the graph is also assumed to contain a globally reachable node.

\section{Concluding remarks}\label{sec:conclusions}

In this work, we have provided a self-contained analysis of smooth K-topical dynamical system. These systems have been proved to have very nice behavior, avoiding periodic trajectories and eventually converging to equilibrium points, if any exist.

We have further investigated the application of these results in the context of multi-agent systems (MASs). K-topicality of the MAS has been shown to be verifiable by only looking at the local interaction rules of the agents. Moreover, standard connectivity conditions on the graph describing the interactions among the agents have been proved to be sufficient to solve the consensus problem in nonlinear K-topical MASs.

This manuscript creates a link bridging monotone dynamical systems theory~\cite{Angeli08} and nonlinear Perron–Frobenius theory~\cite{LemmensNussbaum2012}, by getting rid of the usual notion of strong monotonicity assumption, while focusing on continuously differentiable systems. Moreover, this manuscript paves the way to a variety of lines of research in the context of multi-agent systems which will retrace those investigated for standard linear consensus.

\printbibliography

\end{document}